\documentclass[twocolumn,prl,aps]{revtex4-2}
\usepackage{floatflt}
\usepackage{psfrag}
\usepackage{amsmath}   
\usepackage{amssymb}
\usepackage{amsfonts}
\usepackage{mathrsfs}
\usepackage{graphicx}

\usepackage[dvipsnames]{xcolor}
\definecolor{myblue}{RGB}{38, 157, 206}
\usepackage[colorlinks=true, citecolor= myblue,linkcolor=blue, urlcolor =  myblue]{hyperref}

\begin{document}

\title
{Novel SuperLattice Plasmon Mode in a Grating of 2D Electron Strips}

\author{V.~M.~Muravev$^{a}$, K.~R.~Dzhikirba$^{a}$, A.~A.~Zabolotnykh$^{b}$, P.~A.~Gusikhin$^{a}$, A.~Shuvaev$^{c}$, M.~S.~Ryzhkov$^{c}$, D.~A.~Khudaiberdiev$^{c}$, A.~S.~Astrakhantseva$^{a}$, I.~V.~Kukushkin$^{a}$, A.~Pimenov$^{c}$}
\affiliation{$^a$ Institute of Solid State Physics, RAS, Chernogolovka, 142432 Russia \\
$^b$ Kotelnikov Institute of Radio-engineering and Electronics of the RAS, Mokhovaya 11-7, Moscow 125009, Russia\\
$^c$ Institute of Solid State Physics, Vienna University of Technology, 1040 Vienna, Austria}

\date{\today}

\begin{abstract}
We investigate GaAs/AlGaAs heterostructure membranes with a metasurface made up of a grating of two-dimensional electron system (2DES) strips. Experiments have revealed a strong plasma resonance in the transmission of the metasurface. We have found that a collective effect from the superlattice, along with lateral screening between the strips, leads to the emergence of a new plasmon mode in the metasurface under study. Furthermore, we develop an analytical approach that accurately describes the behavior of the discovered superlattice plasmon mode, providing new insights into the fundamental physics of plasmonic metasurface systems.
\end{abstract}

\maketitle

Plasma excitations in two-dimensional electron systems (2DES) have been the focus of active research for the past 50 years~\cite{Allen:1978, Theis:1980, Heitmann:1986, Koppens:2011, Basov:2016, Lusakowski:2016, Muravev:2020}. This interest stems from potential applications in the field of terahertz (THz) electronics~\cite{Shur:1993, Knap:2002, Shaner:2005, Aizin:2006, Muravev:2012}. Pioneering experiments on plasma waves in two-dimensional semiconductor systems have been carried out using a type of plasmonic metasurface known as a plasmonic crystal~\cite{Allen:1977, Theis:1977, Heitmann:1982}. The plasmonic crystal represents a two-dimensional electron system (2DES) with a periodic gate deposited on the crystal surface over the 2DES. 

Subsequent research focused on plasmon excitations in an array of 2DES disks~\cite{Allen:1983, Leavitt:1986, Heitmann:1990, Dahl:1992, Yan:2012} and strips~\cite{Sarma:1985, Heitmann:1991, Pinczuk:1991, Popov:2010, Wang:2011, Muravev:2022}. The spectrum of two-dimensional (2D) plasmons in such metasurfaces was historically analyzed based on the plasma oscillations in each individual metasurface element~\cite{Kotthaus:1992}. In turn, the plasmon frequency is given by the formula~\cite{Stern:1967, Chaplik:1972}: 
\begin{equation}
    \omega_p (q)=\sqrt{\frac{n_s e^2}{2 \, m^{\ast} \varepsilon_0  \varepsilon} q}  \qquad (q \gg \omega/c).
\label{Plasmon}
\end{equation}
Here, $q$ is the wave vector of the plasmon excitation, $n_s$ and $m^{\ast}$ are the density and the effective mass of the 2DES electrons, while $\varepsilon_0$ and $\varepsilon$ denote the vacuum permittivity and the effective permittivity of the surrounding medium. For the 2DES strip with a width of $w$, the wave vector is typically assumed to be $q \approx 3\pi/4w$~\cite{Quinn:1986, Rudin:1997,Nikitin:2014, Velizhanin:2015, Fogler:2018}. The final frequency of the plasmon excited in the metasurface structure is then determined based on a model of interacting dipoles between the neighboring elements in the lattice.  

It should be noted that the metasurface made up of a grating of 2DES strips shows great potential for terahertz applications. Predictions suggest that such a tunable metasurface could function as an effective THz amplitude and phase modulator, enabling beam steering and wavefront shaping~\cite{Li:2015, Li:2022}. 
The primary factor which distinguishes plasmonic metasurfaces is the significant tunability of the 2DESs. This is achieved by changing the electron density or applying an external magnetic field, allowing for continuous tuning of the properties of the plasmonic metasurface.


In the present paper, we report on terahertz spectroscopy experiments on the GaAs membranes with a grating of strips formed lithographically in a high-mobility 2DES. The developed metasurface demonstrates a strong plasmonic response when the incident radiation polarization is perpendicular to the strips. We find that a superlattice collective effect, together with lateral screening between the strips, leads to the emergence of a new plasmon mode with frequency:
\begin{equation}
    \omega_{sp} =\sqrt{\frac{n_s e^2 }{2 \, m^{\ast} \varepsilon_0  \varepsilon_{\rm eff}} \, \frac{\pi}{p} \, \frac{1}{\ln \left( \dfrac{p}{2 \pi h}\right) + 2}},  
\label{NewPlasmon}
\end{equation}
where $p$ is the metasurface period, $h=p-w$ is the gap between the 2DES strips, which is assumed to be small compared to $p$, and $\varepsilon_{\rm eff} = (1 + \varepsilon_{\rm GaAs})/2$ is the effective dielectric permittivity. The developed theory~(\ref{NewPlasmon}) has proved to be in good agreement with experimental data. Note that the frequency~(\ref{NewPlasmon}) approaches zero as the gap $h$ closes. This is the distinguishing feature of the novel superlattice plasmon mode. The obtained results establish the scientific grounds for constructing plasmonic metasurface elements for the development of terahertz systems.

\begin{figure}[!t]
    \centering
    \includegraphics[width=\linewidth]{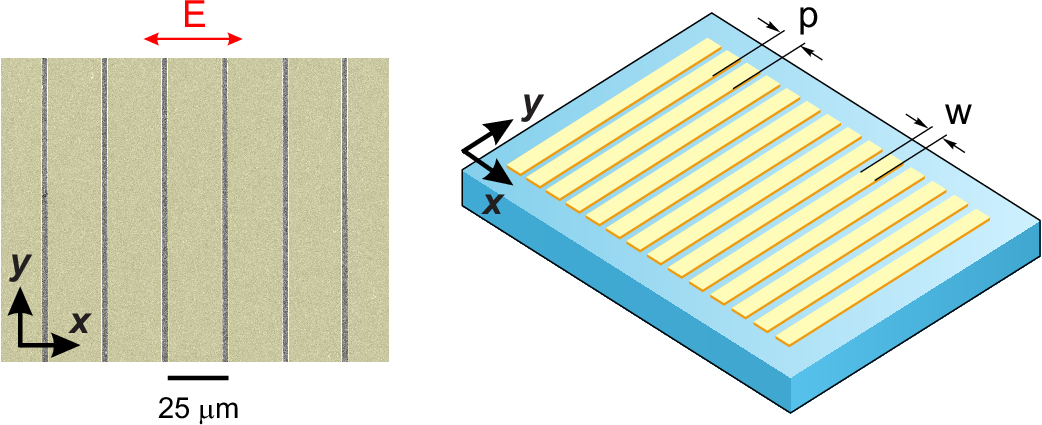}
    \caption{(Left) E-beam photo of the sample comprising a grating of isolated 2DES strips. The area of the 2DES is highlighted in yellow. Each 2DES strip has a width of $w = 23$~\textmu{}m, and the period is $p = 25$~\textmu{}m. (Right) Schematic of the plasmonic metasurface device. The two-dimensional electron system is represented in yellow, while the GaAs substrate is shown in blue.
    The coordinate system is provided in both panels to indicate the orientation of the structures and to facilitate comparison between the micrograph and the schematic.}
    \label{1}
\end{figure}

The samples were made from the industrial Al$_{0.3}$Ga$_{0.7}$As/GaAs/Al$_{0.3}$Ga$_{0.7}$As heterostructures grown by the molecular beam epitaxy. The wafer has a $20$~nm  wide quantum well grown at a distance of $210$~nm below the crystal surface. The quantum well hosts the 2DES with the density $n_s = 9.3 \times 10^{11}$~cm$^{-2}$ and electron mobility $\mu = 10^5$~cm$^2$/Vs at $T=5$~K. In computations, we take an effective mass of electrons $m^{\ast} = 0.071 \, m_0$ due to the nonparabolicity of the electron conduction band occurring at such high 2D electron density~\cite{Francisco:1988, Ekenberg:1989}. The semiconductor wafer was cut into  $1 \times 1$~cm$^2$ samples. Then, we lithographically fabricated a grating of 2DES isolated strips (Fig.~\ref{1}). For eight samples, the metasurface period is fixed $p=25$~\textmu{}m and the strip width varies from $w=13$~\textmu{}m to $23.7$~\textmu{}m. Details of the samples geometry are given in the Supplemental Material~\cite{Supplemental}. All samples were etched from the back side with a citric acid solution to achieve a uniform membrane whose thickness varied for different samples from $35$ to $45$~\textmu{}m~\cite{Astrakhantseva:2022}. The sample was mounted at the end of the sample holder, which was inserted into the cryostat at the center of the superconducting coil. The cryostat was equipped with two mylar windows coplanar to the surface of the sample. The phase/intensity of the electromagnetic wave that passed through the sample was investigated using a Mach-Zehnder interferometer~\cite{Kozlov:1998}. A set of backward-wave oscillators (BWO) has been used as a source of radiation in the frequency range of interest $f=(40-500)$~GHz. The electromagnetic radiation was polarized perpendicularly to the 2DES strips. The transmittance through the sample was measured using a He-cooled bolometer. All experiments were conducted at a base sample temperature of $5$~K.

\begin{figure}[!t]
    \centering
    \includegraphics[width=\linewidth]{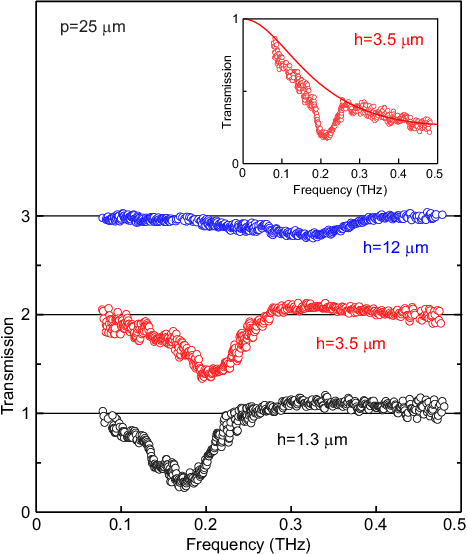}
    \caption{The normalized transmission spectra of three samples with different strip spacings, specifically $h = 12$, $3.5$, and $1.3$~\textmu{}m. All three samples share the same period, $p = 25$~\textmu{}m. For clarity, the curves have been vertically shifted by $1$ unit. The inset shows the transmission spectrum of the $h = 3.5$~\textmu{}m sample without normalization. The solid red line represents the Fabry-P\'erot function (\ref{FP}) for the bare dielectric membrane with no 2DES.}
    \label{2}
\end{figure}

The inset to Fig.~\ref{2} shows the transmission spectrum recorded for a metasurface period of $p=25$~\textmu{}m and a strip width of $w=21.5$~\textmu{}m. A clear plasmon resonance is observed at $f_p=0.2$~THz, superimposed on the Fabry-P\'erot transmission function of the bare dielectric membrane~\cite{Gruner}:
\begin{equation}
    T_{\rm FP} = \frac{1}{1 + \dfrac{1}{4} \left( \sqrt{\varepsilon} - \dfrac{1}{\sqrt{\varepsilon}}   \right)^2 \sin^2 \left( \dfrac{\sqrt{\varepsilon} \, \omega d}{c} \right)},
    \label{FP}
\end{equation}
where $\varepsilon = 12.8$ is the dielectric constant of GaAs, and thickness of the membrane $d=40$~\textmu{}m. 

To eliminate the Fabry-P\'erot background and spurious resonances of the quasi-optical path, we divide the transmission at $B=0$~T by the transmission through the sample at $B=7$~T. At such a strong magnetic field, when $\omega_c \tau =(eB/m^{\ast}) \tau \gg 1$ and $\omega_c \gg \omega$, the motion of the 2D electrons in the quantum well plane is quenched in cyclotron orbits. As a consequence, the 2DES  does not affect the transmittance~\cite{Muravev:2023PRA}. The resultant normalized transmission spectra recorded for three samples with different 2DES strip widths $w=13, 21.5$ and $23.7$~\textmu{}m but fixed metasurface period $p=25$~\textmu{}m are shown in Fig.~\ref{2}. The plasmon resonance moves to a higher frequency as the gap between the strips is increased. 

\begin{figure}[!t]
    \centering
    \includegraphics[width=\linewidth]{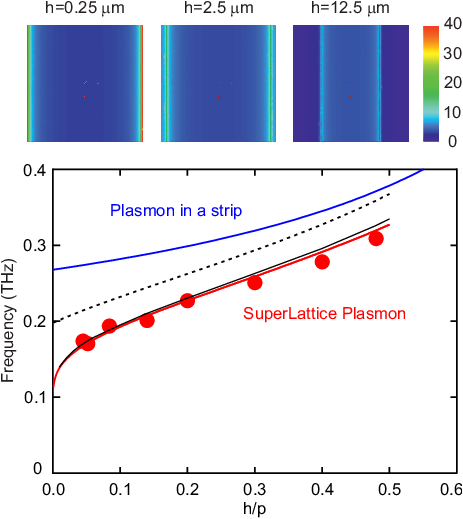}
    \caption{Dependence of the plasmon frequency on the $h/p$. The period of the plasmonic metasurface was maintained at a fixed value of $p = 25$~\textmu{}m. The solid red line represents a theoretical prediction~(\ref{NewPlasmon}), while the blue line presents the standard 2D plasmon frequency (\ref{Plasmon}) in a single strip with wave vector $q=3 \pi/4w$. The solid black line shows the result of simulation in the HFSS 3D high-frequency simulation software. The dotted curve in Fig.~\ref{3} depicts the result for the analytical prediction from~\cite{Velizhanin:2015}. Top inset displays the color maps of field-$|E|$ distribution at the plasmon resonance for $h=0.25$, $2.5$, and $12.5$~\textmu{}m ($p = 25$~\textmu{}m). The 2DES strip is located at the center of each color map.}
    \label{3}
\end{figure}

However, the experimentally obtained values of the plasmon frequency differ significantly from those predicted by the standard dispersion equation (\ref{Plasmon}). Indeed, Fig.~\ref{3} shows how the plasmon frequency depends on the gap between the 2DES strips $h/p=(p-w)/p$. This dependence was measured across eight samples, all with the same period of $p=25$~\textmu{}m. The solid blue line in the Figure represents the theoretical prediction based on equation (\ref{Plasmon}) with $q=3 \pi/4w$. It is clear that there is a significant discrepancy between this theoretical prediction and the experimental results, even with significant spacing between the individual strips. To better understand the situation underway, we have developed a theoretical framework to describe the behavior of the plasmonic metasurface under consideration.

The theoretical analysis is based on the solution of Maxwell's equations and the determination of the response of the system to the external radiation incident normally on the metasurface. Overall, the approach follows to that applied in Refs.~\cite{Mikhailov:1998, khisameeva_prr_2025} devoted to the electromagnetic response of plasmonic crystals. 

Since the system is periodic the electric and magnetic fields can be decomposed into spatial series as follows:
\begin{equation}
    \begin{split}
    & {\bf E}(x,z,t)=\sum_{k=-\infty}^{\infty} {\bf E}^{k}(z) e^{iq_k x-i\omega t};\\ 
    & {\bf H}(x,z,t)=\sum_{k=-\infty}^{\infty} {\bf H}^{k}(z) e^{iq_k x-i\omega t},
    \end{split}
\end{equation}
where $k$ and $q_k=2 \pi k/p$ are the number of the harmonic and the corresponding wave vector, $x$-axis is directed across the strips, while $y$- and $z$-axes are directed respectively along the strips and perpendicular to the metasurface (Fig.~\ref{1}). 

In the absence of an external magnetic field, and as the electric field of the incident radiation polarized in the $x$-direction, only the transverse magnetic (TM) modes with the field components $(E_x,H_y,E_z)$ are excited in the system. The electromagnetic response is described by the ratio of the self-consistent $x$-component of the electric field arising in a gap between the strips to the amplitude of $E_x$-component in the incident radiation.  The maxima of this ratio correspond to the excitation of TM-modes in the system. The detailed derivation of the response function can be found in the Supplemental Material~\cite{Supplemental}, while the main points and approximations are discussed below.

Firstly, we note that the period of the structure, the distance between the center of the quantum well and the crystal surface, $d=220$~nm, and the thickness of the membrane (substrate), $D=(35-45)\,\mu$m, are small compared to the wavelength of the radiation. At a frequency of $f = 200$ GHz, the wavelength is $\lambda = c/ f\sqrt{\varepsilon_{\rm GaAs}}\approx 0.42$~mm.  This indicates that electromagnetic retardation effects can be neglected when describing the TM-modes with finite values of $q_k$ (where $k = 1, 2, \ldots$). Essentially, these modes are quasi-electrostatic plasmons that occur in an array of 2D strips~\cite{Quinn:1986}. However, the influence of electromagnetic retardation persists in the $k = 0$ spatial harmonic, which accounts for the scattered electromagnetic waves and leads to the radiative broadening of the plasma resonances.

Secondly, since in the experiment under consideration the distance $h$ is relatively small --- specifically, the ratio $h/p$ does no exceed $0.5$ ---  it is natural to derive and solve the equation for the electric field $E_x$ in the gap between the strips of the metasurface. Next, to find the frequencies of plasma resonances explicitly, we use a model function for the desired electric field, namely, we take the electric field to be uniform within the gap. This model can be treated as the use of one function in the expansion of the field into a series of polynomials and, in principle, a larger number of polynomials can be taken into account. Nevertheless, the uniform field model provides results that already agree well with experimental data.

Under the assumption that the electron relaxation rate due to collisions with impurities and phonons as well as the radiative decay rate of plasma modes are both significantly less than the resonant frequencies, the latter can be determined as the zeros of the denominator of the response function~\cite{Supplemental}. With above approximations in place, resonant plasmon frequencies can be found from the following equation:
\begin{equation}
\label{T:den}
    1+ 2 \sum_{k=1}^{+\infty} \frac{1}{1-\omega_p^2 k\varkappa_k/\omega^2}\left(\frac{\sin(\pi k h/p)}{\pi k h/p}\right)^2=0,
\end{equation}
where $\omega_p$ is defined by Eq.~(\ref{Plasmon}) with $q=2 \pi /p$ and $\varepsilon=\varepsilon_{\rm GaAs}$. The coefficient $\varkappa_k =1+ (\varepsilon-1)e^{-2q_k  d}/(\varepsilon+1)$ results from the finite thickness of the cap layer $d$. Throughout derivation of formula (\ref{T:den}), the high-frequency Drude model for conductivity, $\sigma(\omega)=e^2 n_s/(-i\omega m^{\ast})$, is assumed. 

It is important to note that Eq.~(\ref{T:den}) describes all resonances that can be excited by a plane wave incident normally on the metasurface. In principle, the frequencies of the plasmon resonances can be determined numerically using Eq.~(\ref{T:den}). However, in the analysis below, we will focus solely on the analytical investigation of the fundamental plasmon resonance. We expect that the frequency of this resonance is low and the condition $\omega^2 \ll \omega_p^2$ is fulfilled. Under this condition, we neglect one in the denominator of all terms in the sum (\ref{T:den}), except for the term with $k=1$. Additionally, we take that strips are situated just at the surface of the substrate and neglect the value of the finite cap layer thickness $d$. Accordingly, we substitute $\varkappa_k$ in (\ref{T:den}) with $\bar{\varkappa} =(\varepsilon_{\rm GaAs}+1)/2$, which corresponds to the average dielectric permittivity. Then, under the condition that $h \ll p$, the sum in the left-hand side of Eq.~(\ref{T:den}) can be found as follows:
\begin{equation}
\label{log_factor}
    \sum_{k=1}^{+\infty} \frac{\sin^2(\pi k h /p)}{k^3 (\pi h/p)^2 }\approx \ln\left(\frac{p}{2 \pi h}\right)+\frac{3}{2}.
\end{equation}
After that we arrive at the final expression (\ref{NewPlasmon}) for the frequency of the fundamental superlattice plasmon mode. It is the logarithmic factor~(\ref{log_factor}) that represents the mutual capacitance between the strips leading to zero frequency of the superlattice plasmon mode as the gap $h$ closes. There is a remarkable agreement between the analytical formula (\ref{NewPlasmon})
displayed in Fig.~\ref{3} by a solid red curve and the experimental data. Additional experiments conducted on a series of samples with a fixed gap between the 2DES strips, ($h=2$~\textmu{}m), and superlattice periods $p=25, 20$ and $14.5$~\textmu{}m validate that the plasmon frequency is governed by the full set of metasurface parameters and not the duty cycle, $h/p$, alone (see Supplemental Material~\cite{Supplemental}).   

To further validate our theoretical model, we performed full-wave numerical simulations using the HFSS finite-element solver (3D High Frequency Simulation Software). The fabricated arrays comprise approximately 400 periods, allowing us to model the structure as an infinite periodic grating by applying periodic boundary conditions. We employed automatic adaptive meshing to ensure convergence of the eigenfrequency and field profiles. The simulated superlattice plasmon frequency is shown by the solid black line in Fig.~\ref{3}, exhibiting excellent agreement with both the proposed model and the experimental data. The simulations also elucidate the physical character of the superlattice mode. The top inset of Fig.~\ref{3} presents color maps of the electric-field magnitude, $|E|$, at resonance for strip separations $h=0.25$, $2.5$, and $12.5$~\textmu{}m. For large separations, the field is concentrated predominantly within the body of the 2DES strip. In contrast, as the gap narrows, the field localizes in the inter-strip region with pronounced enhancement.

It is also instructive to consider the limit of a vanishing gap between adjacent 2DES strips. In this case, the frequency of the fundamental plasmon mode predicted by Eq.~(\ref{NewPlasmon}) tends to zero. At the same time, the structure continuously evolves into a uniform, ungated 2D electron channel, whose plasmon spectrum is governed by the ordinary dispersion relation, Eq.~(\ref{Plasmon}). These two statements are not contradictory. As shown in the Supplemental Material~\cite{Supplemental}, within our theoretical framework the higher-order plasmon resonances approach the ordinary 2D plasmon dispersion, Eq.~(\ref{Plasmon}), with wave vectors $q = (2\pi/p) \times N$ ($N=1,2,\ldots$) in the limit of a disappearing inter-strip gap.

We should note that several attempts have been made to analytically describe plasmons in the metasurface under study~\cite{Khavasi:2014, Velizhanin:2015}. However, these theories start with a trial function for the field (or current) within the strips and then extend their consideration to the lattice of strips. This approach overlooks the effects of the superlattice. For instance, the dotted curve in Fig.~\ref{3} represents the result for the analytical prediction from~\cite{Velizhanin:2015}. The agreement with experimental results is not satisfactory.

\begin{figure}[!t]
    \centering
    \includegraphics[width=\linewidth]{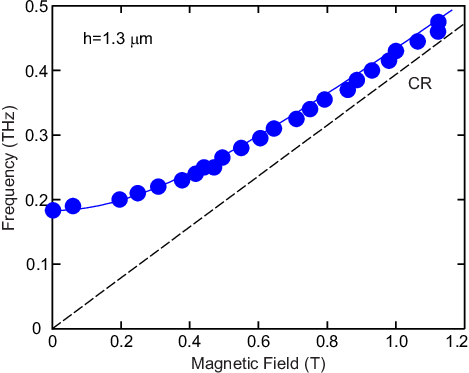}
    \caption{Magnetodispersion of the observed plasma resonance. The dashed line denotes the cyclotron resonance $\omega_c = e B/m^{\ast}$. The solid line represents the hybrid magnetoplasma law $\omega^2 = \omega_{sp}^2 + \omega_c^2$. The metasurface strip width is $w = 23.7$~\textmu{}m, with a period of $p = 25$~\textmu{}m.}
    \label{4}
\end{figure}

To further explore the physical nature of the observed superlattice plasmon mode, we measured the frequency position of the plasmon resonance as a function of the magnetic field applied perpendicular to the 2DES plane (Fig.~\ref{4}). The experiment was conducted with a metasurface having a gap between the 2DES strips of $h = 1.3$~\textmu{}m. The resonance starts at a plasmon frequency $\omega_{sp}/2 \pi = 175$~GHz and rapidly saturates at the cyclotron resonance frequency, given by $\omega_c = e B/m^{\ast}$ (dashed line in Fig.~\ref{4}). The resonance frequency follows the law $\omega^2 = \omega_{sp}^2 + \omega_c^2$, depicted by solid line in Fig.~\ref{4}. The magnetodispersion in Fig.~\ref{4} is typical for quasi-electrostatic magnetoplasma excitations~\cite{Theis:1977}. The revealed magnetodispersion is also briefly discussed in Supplemental Material~\cite{Supplemental}.

In conclusion, we have investigated the terahertz response of a semiconductor membrane with a periodic grating of strip-shaped 2DES elements. When the wavelength of the incident radiation is much larger than the grating period, the device behaves like a plasmonic metasurface whose properties can be engineered by the geometry of the strip elements. We discovered that a collective effect from the superlattice, combined with lateral screening between the strips, gives rise to the emergence of a new superlattice plasmon mode. This mode arises from the superposition of standing plasma waves across all wave-vectors of the metasurface's reciprocal lattice. We developed an analytical expression for the frequency of the mode, which shows remarkable agreement with experimental results. The distinguishing feature of the novel mode is that its frequency approaches zero as the gap between the strips closes. The obtained results establish the scientific foundation for the development of plasmonic metasurface components for terahertz electronics.

\begin{acknowledgments}
The authors gratefully acknowledge the financial support of the Russian Science Foundation (Grant No.~19-72-30003) and the TU Wien Bibliothek for financial support through its Open Access Funding Program.

\end{acknowledgments}

\bibliography{main}

\end{document}


\title{Supplementary Material for\\ ``Novel SuperLattice Plasmon Mode in a Grating of 2D Electron Strips''}

\author{V.~M.~Muravev$^{a}$, K.~R.~Dzhikirba$^{a}$, A.~A.~Zabolotnykh$^{b}$, P.~A.~Gusikhin$^{a}$, A.~Shuvaev$^{c}$, M.~S.~Ryzhkov$^{c}$, D.~A.~Khudaiberdiev$^{c}$, A.~S.~Astrakhantseva$^{a}$, I.~V.~Kukushkin$^{a}$, A.~Pimenov$^{c}$}
\affiliation{$^a$ Institute of Solid State Physics, RAS, Chernogolovka, 142432 Russia \\
$^b$ Kotelnikov Institute of Radio-engineering and Electronics of the RAS, Mokhovaya 11-7, Moscow 125009, Russia\\
$^c$ Institute of Solid State Physics, Vienna University of Technology, 1040 Vienna, Austria}

\date{\today}
\maketitle

\section{\textrm{I}. Detailed description of all measured samples}

The information on all the samples under study is presented in the table. The samples with an active area of $8 \times 8$~mm were fabricated using standard photolithography.

\begin{table}[h]
    \centering
    \begin{tabular}{|c|c|c|c|c|c|c|c|c|c|c|c|c|c|c|c|}
        \hline
        \textbf{Structure parameters} & \textbf{S1} & \textbf{S2} & \textbf{S3} & \textbf{S4} & \textbf{S5} & \textbf{S6} & \textbf{S7} & \textbf{S8} \\ 
        \hline
         $h$ ($\mu$m) & $1.12$  & $1.3$ & $2.1$ & $3.5$ & $5$ & $7.5$  & $10$ & $12$ \\
        \hline
         $h/p$ & \quad $0.045$ \quad & \quad $0.052$ \quad & \quad $0.084$ \quad & \quad $0.14$ \quad & \quad $0.2$ \quad & \quad $0.3$ \quad & \quad $0.4$ \quad & \quad $0.48$ \quad \\
        \hline
    \end{tabular}
    \caption{Parameters of the samples with a superlattice period of $p = 25$~\textmu{}m.}
\end{table}

\section{\textrm{II}. Analytical description of plasmon modes in the grating of 2D strips}
\begin{figure}[h]
    \centering
    \includegraphics[width=0.45\linewidth]{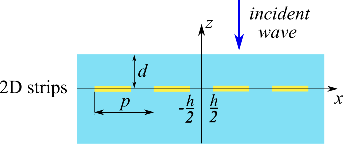}
    \caption{Schematic of the structure under study.}
    \label{STheor}
\end{figure}

Consider 2D stripes situated at the plane $z=0$ as shown in Fig.~{\ref{STheor}}. Strips have width $w$ in the $x$-direction, are infinite along the $y$-axis and $\delta$-thin in $z$-direction. The dielectric permittivity $\varepsilon(z)$ equals unity at $z>d$ and $\varepsilon$ at $z<d$, which corresponds to GaAs substrate of infinite thickness. The external electromagnetic wave falls on the system against $z$-axis. The electric field of the incident wave is polarized in the $x$-direction and, consequently, only so-called transverse magnetic (TM) modes with field components $(E_x, H_y, E_z)$ arise in the system. The situation changes for the system placed in the constant perpendicular magnetic field, the issue is discussed in the next section.

To determine the response of the system, we follow the usual approach \cite{Mikhailov:1998,Mikhailov:2005,khisameeva_prr_2025} based on the solution of the Maxwell's equations, which in CGS units appear as follows:
\begin{equation}
\label{Supp:Maxeq}
    rot { \bf H}=\frac{\varepsilon(z)}{c} \frac{\partial {\bf E} }{\partial t } +\frac{4 \pi }{c} {\bf j}, \quad rot {\bf E}=-\frac{1}{c} \frac{\partial {\bf H} }{\partial t },
\end{equation}
where $c$ is the speed of light in vacuum and ${\bf j}={\bf j}^{2D}(x,t)\delta(z)$ is the charge density current in 2D strips, ${\bf j}$ is directed along $x$-axis.

Since we look for periodic in $x$-directions solutions, we introduce the following Fourier series for the sought functions:
\begin{equation}
\label{Supp:form}
    E_{x,z}(x,z,t)=\sum_{k=-\infty}^{\infty} E_{x,z}^{k}(z) e^{iq_k x-i\omega t}, \, \,
    H_{y}(x,z,t)=\sum_{k=-\infty}^{\infty} H_{y}^{k}(z) e^{iq_k x-i\omega t}, \,\, j_{x}^{2D}(x,t)=\sum_{k=-\infty}^{\infty} j_{x}^{2D,k}e^{iq_k x-i\omega t},
\end{equation}
where $q_k=2\pi k/p$ is the $k$-th space harmonic of the grating, $p$ is the period of the structure, and $\omega$ is the frequency of the wave.

We introduce relations in (\ref{Supp:form}) into Eqs.~(\ref{Supp:Maxeq}), exclude fields $H_{y}^k(z)$ and $E_z^k(z)$ as follows:
\begin{equation}
    H_y^k= \frac{c}{i \omega} \left( \partial_z E_x^k -i q_k E_z^k\right), \quad
    E_z^k=\frac{-i q_k}{q_k^2-\omega^2 \varepsilon(z)/c^2} \partial_z E_x^k, 
\end{equation}
and obtain the connection between $E_x^k(z)$ and $j_x^{2D,k}$:
\begin{equation}
\label{Supp:waveX}
    \partial_z \frac{\varepsilon(z)}{q_k^2-\omega^2 \varepsilon(z)/c^2} \partial_z E_x^k(z) - \varepsilon(z) E_x^k(z) = \frac{4 \pi}{ -i \omega} j_x^{2D,k}\delta(z),
\end{equation}
The above equation describe TM electromagnetic modes in the system taking into account the inhomogeneity of the dielectric permittivity $\varepsilon(z)$ along the $z$-axis and the presence of currents in 2D stripes.

Using Eq.~(\ref{Supp:waveX}) we consider the scattering of the plane electromagnetic wave $E_0 e^{-i\omega z/c -i\omega t}$ incident on the system against the $z$-axis as shown in Fig.~\ref{STheor}. Outside the interfaces $z=0$ and $z=d$ the electric field of the waves has the form as follows:
\begin{equation}
\label{Supp:formX}
E_x^k(z)=
\begin{cases}
     C_{1}^k e^{-\beta^0_k z} +E_0 e^{-i\omega z/c} \delta_{k0}, & z\geq d; \\
     C_{2}^k e^{\beta_k z}+ C_{3}^k e^{-\beta_k z}, & d>z>0; \\
     C_{4}^k e^{\beta_k z}, & z\le 0;
\end{cases}
\end{equation}
where $\delta_{k0}$ is the Kronecker delta, which equals one at $k=0$ and zero at $k \neq 0$, while coefficients $\beta_k=\sqrt{q_k^2-\omega^2 \varepsilon/c^2}$ and $\beta_k^0=\sqrt{q_k^2-\omega^2/c^2}$ determine how field $E_x^k(z)$ extends into dielectric medium and vacuum correspondingly. Note that $\beta_k$ and $\beta_k^0$ can be real (if $q_k>\omega \sqrt{\varepsilon}/c$ or $q_k>\omega/c$) or imaginary (if the opposite inequalities take place). Real values of $\beta_k$ and $\beta_k^0$ correspond to the waves localized in the $z$-direction, i.e. evanescent waves. Imaginary values of $\beta_k$ and $\beta_k^0$ describe propagating waves and the sign of the imaginary part should be {\it negative}, which corresponds to the waves outgoing from the system; for instance, since here and below we assume $\omega>0$ then at $k=0$ we obtain $\beta^0_{k=0}=-i\omega/c$ and $\beta_{k=0}=-i\omega \sqrt{\varepsilon}/c$.

To find the solution, along with Eq.~(\ref{Supp:formX}), we need to impose boundary conditions for the electric field $E_x^k(z)$ at $z=0$ and $z=d$. The first condition is the continuity of $E_x^k(z)$, since it is tangent to the surfaces $z=0$ and $z=d$. The second boundary condition can be derived via the integration of Eq.~(\ref{Supp:waveX}) near $z=0$ and $z=d$:
\begin{equation}
\label{Supp:BC}
    \frac{\varepsilon(z)}{q_k^2-\omega^2 \varepsilon(z)/c^2} \partial_z E_x^k(z) |^{z=+0}_{z=-0}= \frac{4 \pi}{ -i \omega} j_x^{2D, k}, \quad \frac{\varepsilon(z)}{q_k^2-\omega^2 \varepsilon(z)/c^2} \partial_z E_x^k(z) |^{z=d+0}_{z=d-0}=0.
\end{equation}

Using the boundary conditions at $z=0$ and the continuity of $E_x^k(z)$ at $z=d$ one can find
\begin{equation}
\label{Supp:Coeffs}
    C^k_2=C^k_4+\frac{2 \pi \beta_k}{-i\omega \varepsilon} j_x^{2D,k}, \quad
    C^k_3=\frac{2 \pi \beta_k}{i\omega \varepsilon} j_x^{2D,k}, \quad 
    C^k_1 e^{-\beta^0_k d}=-E_0 e^{-i\omega d/c} \delta_{k0} +C^k_2 e^{\beta_k d} +C^k_3 e^{-\beta_k d}.
\end{equation}
Substituting the above coefficients into the second relation in~(\ref{Supp:BC}) and taking into account that $E_x^k(z=0)=C^k_4$ results in the desired equation:
\begin{equation}
\label{Supp:eqsE}
    E_x^k(z=0)=\frac{2}{\sqrt{\varepsilon}+1}E_0 e^{i\omega d(\sqrt{\varepsilon}-1)/c}\delta_{k0} -\frac{2 \pi \beta_k}{-i\omega \varepsilon}\left(1+ \frac{\varepsilon/\beta_k-1/\beta_k^0}{\varepsilon/\beta_k+1/\beta_k^0} e^{-2\beta_k d}\right) j_x^{2D,k},
\end{equation}
where
\begin{equation}
\label{Supp:jx}
    j_x^{2D,k}=\int_{strip} \frac{dx'}{p} j^{2D}_x(x') e^{-iq_k x'}.
\end{equation}

Finally, to find the response of the system the above equations should be supplemented by the Ohm's law. The latter in its local form appears as $j_x^{2D}(x)=\sigma E_x(x,z=0)$, where $x$ lies {\it inside} a strip and $\sigma=\sigma(\omega)$ is the dynamical (optical) conductivity of a strip. However, since in the experiment the distance between the strips, $h$, is rather small, namely, the maximal value of $h/p$ is near $0.5$, it is natural to rewrite Eq.~({\ref{Supp:eqsE}}) for the electric field in the gap between the strips, excluding the current in the strips.

The current in Eq.~(\ref{Supp:jx}) can be written as follows:
\begin{multline}
\label{Supp:jx2}
    j_x^{2D,k}= \int_{-p/2}^{p/2} \frac{dx'}{p} \sigma E_x(x',z=0) e^{-iq_k x'}- \int_{-h/2}^{h/2} \frac{dx'}{p} \sigma E_x(x',z=0) e^{-iq_k x'}=\\
    \sigma E_x^k(z=0)-\sigma \int_{-h/2}^{h/2} \frac{dx'}{p} E_x(x',z=0) e^{-iq_k x'}, 
\end{multline}
where the conductivity $\sigma$ does not depend on coordinate $x$.

We substitute the above expression for $j_x^{2D,k}$ into Eq.~(\ref{Supp:eqsE}) and express $E_x^k(z=0)$. After that, the following equation for $E_x(x,z=0)$ can be achieved:
\begin{equation}
\label{Supp:Efull}
     E_x(x,z=0)=\frac{2}{\sqrt{\varepsilon}+1} \frac{E_0  e^{i\omega d (\sqrt{\varepsilon}-1)/c}}{1+\frac{2 \pi \sigma}{c \sqrt{\varepsilon}}\left(1+\frac{\sqrt{\varepsilon}-1}{\sqrt{\varepsilon}+1}e^{2i\omega \sqrt{\varepsilon} d/c}\right)}+
     \sum_{k=-\infty}^{+\infty} \frac{\frac{2 \pi \sigma \beta_k}{-i\omega \varepsilon}\widetilde\varkappa_k}{1+\frac{2 \pi \sigma \beta_k}{-i\omega \varepsilon}\widetilde\varkappa_k}\int_{-h/2}^{h/2} \frac{dx'}{p} E_x(x',z=0) e^{iq_k (x-x')},
\end{equation}
where
\begin{equation}
    \widetilde\varkappa_k=1+ \frac{\varepsilon/\beta_k-1/\beta_k^0}{\varepsilon/\beta_k+1/\beta_k^0} e^{-2\beta_k  d}.
\end{equation}

Finally, Eq.~(\ref{Supp:Efull}) can be rewritten in the following form:
\begin{multline}
\label{Supp:Efin}
     E_x(x,z=0)=\frac{2}{\sqrt{\varepsilon}+1} \frac{E_0  e^{i\omega d(\sqrt{\varepsilon}-1)/c}}{1+\frac{2 \pi \sigma}{c \sqrt{\varepsilon}}\left(1+\frac{\sqrt{\varepsilon}-1}{\sqrt{\varepsilon}+1}e^{2i\omega \sqrt{\varepsilon} d/c}\right)}+
     E_x(x,z=0)\left(\theta(x+h/2)-\theta(x-h/2)\right)-\\
     \sum_{k=-\infty}^{+\infty} \frac{1}{1+\frac{2 \pi \sigma \beta_k}{-i\omega \varepsilon}\widetilde\varkappa_k}\int_{-h/2}^{h/2} \frac{dx'}{p} E_x(x',z=0) e^{iq_k (x-x')},
\end{multline}
Note that the above equations, as well as Eq.~(\ref{Supp:eqsE}), follow directly from Maxwell's equations and the local Ohm's law without any approximations.

Now, we make a few assumptions. First of all, in the experiment the period of the structure (as well as the distance $d =220$\,nm) is much less than the wavelength of the electromagnetic radiation. The latter at frequency $f=200$~GHz in GaAs with $\varepsilon_{\rm GaAs}=12.8$ is $c/f\sqrt{\varepsilon_{\rm GaAs}} \approx 420$\,\textmu{}m, while the period $p=25$\,\textmu{}m. Therefore, $2\pi/p \gg \omega/c, \, \omega \sqrt{\varepsilon}/c$ and, consequently, we take $\beta_k=\beta^0_k=|q_k|$ for $k \neq 0$, whereas $\beta_{k=0}=-i\omega\sqrt{\varepsilon}$ and $\beta_{k=0}^0=-i\omega/c$. In essence, in this limit waves with $k \geq 1$ correspond to quasi-electrostatic 2D plasmons, which are localized in $z$-direction, and only scattered waves with $k=0$, i.e. uniform in $x$-direction, can freely propagate along and against $z$-axis, see Eq.~(\ref{Supp:formX}).

Secondly, to find the desired resonant frequencies explicitly, we use the model function for the electric field $E_x(x,z=0)$ in a gap. This electric field is qualitatively due to charges of different signs induced near edges of the adjacent strips, so the electric field inside a gap at least has no zeros and is directed from one strip to another. That is why, as a model, we take the electric field to be uniform in a gap. Note that this model can be considered as the use of one (constant) function in the expansion of the field $E_x(x,z=0)$ into a series of polynomials on the interval $(-h/2,h/2)$ and, if necessary, a larger number of polynomials can be taken into account. However, as will be seen below, the model of uniform field in a gap already gives results that are in good agreement with experimental data.

Using the first approximation and taking that $E_x(x,z=0)$ is the even function, Eq.~(\ref{Supp:Efin}) becomes as follows:
\begin{multline}
\label{Supp:E2}
     E_x(x,z=0)=\frac{2E_0}{\sqrt{\varepsilon}+1+4 \pi \sigma/c }+ E_x(x,z=0)\left(\theta(x+h/2)-\theta(x-h/2)\right)-\frac{1}{1+\frac{4 \pi \sigma}{c(\sqrt{\varepsilon}+1)} }\int_{-h/2}^{h/2} \frac{dx'}{p} E_x(x',z=0)- \\
     2 \sum_{k=1}^{+\infty} \frac{1}{1+\frac{2 \pi \sigma q_k}{-i\omega \varepsilon}\varkappa_k}\int_{-h/2}^{h/2} \frac{dx'}{p} E_x(x',z=0) e^{iq_k (x-x')}, \quad \text{where} \,\,  \varkappa_k=1+ \frac{\varepsilon-1}{\varepsilon+1} e^{-2q_k  d}. 
\end{multline}

Now we use the second approximation, namely, we consider the above equation inside a gap, i.e. at $-h/2<x<h/2$, substitute $E_x(x,z=0)=E_g$, and integrate the obtained relation from $-h/2$ to $h/2$. Then the response of the system can be found as follows:
\begin{equation}
\label{Supp:resp}
    \frac{E_g}{E_0}=\frac{2}{\sqrt{\varepsilon}+1+4 \pi \sigma/c} \frac{p/h}{\left(1+\frac{4\pi \sigma}{c(\sqrt{\varepsilon}+1)}\right)^{-1}+2\left(\frac{p}{h} \right)^2 \sum_{k=1}^{+\infty} \left(1+\frac{2 \pi \sigma q_k}{-i\omega \varepsilon}\varkappa_k\right)^{-1}\left(\frac{\sin(\pi k h/p)}{\pi k}\right)^2}.
\end{equation}
The first factor in the above equation is the smooth function of frequency $\omega$ and it is the denominator of the second factor that determines the resonant frequencies as well as line widths.

To proceed we use the Drude model for the conductivity, which appears as follows: $\sigma(\omega)=e^2 n_s /m^{\ast} (-i\omega +1/\tau)$, where $n_s$, $m^{\ast}$, and $\tau$ are respectively 2D electron concentration, the effective mass, and the relaxation time. In the experiment, the relaxation time, estimated from the electron mobility $\mu=e \tau /m^{\ast} =10^5$\, cm${^2}$/V$\cdot$ s, is $\tau=4$\,ps, where we take $m^{\ast}=0.071m_0$ with $m_0$ being the mass of a free electron (this value emerges due to the influence of nonparabolicity of the electron conduction band at a sufficiently high concentration of $n_s =9.3 \times 10^{11}$ cm$^{-2}$~\cite{Francisco:1988, Ekenberg:1989}). Thus, the relation $\omega \tau \gg 1$ takes place for typical experimental frequency of 200 GHz. With the use of the Drude model in the limit $\omega \tau \gg 1$, the denominator in~(\ref{Supp:resp}) becomes as follows:
\begin{equation}
\label{Supp:den}
    \frac{1}{1+i\Gamma/\omega}+ 2\left(\frac{p}{h} \right)^2 \sum_{k=1}^{+\infty} \frac{1}{1-\omega_p^2 k\varkappa_k/\omega^2}\left(\frac{\sin(\pi k h/p)}{\pi k}\right)^2, \quad \text{where} \,\, \Gamma=\frac{4\pi e^2 n_s}{m^{\ast}c(\sqrt{\varepsilon}+1)}, \,\, \omega_p=\sqrt{\frac{2\pi e^2 n_s}{m^{\ast}\varepsilon}\frac{2 \pi}{p}}
\end{equation}
is the frequency of 2D plasmons with the wave vector $2 \pi/p$. The factor $\Gamma$ is related with the radiative contribution to the line width. The value of $\Gamma/(2\pi)$ in the case of our system is near 48 GHz, which is less than the typical frequency of 200 GHz. So, while determining the resonant frequencies we neglect $\Gamma/\omega$ in (\ref{Supp:den}).

Importantly, that the obtained relations (\ref{Supp:resp}) and (\ref{Supp:den}) describes all resonances that can be exited by plane waves falling normally on the system. Before we proceed it is interesting to point out the vanishing of the resonance response of the system due to zeros of $\sin(\pi k h/p)$ in Eq.~(\ref{Supp:den}). Indeed, at $\sin(\pi k h/p)=0$, i.\,e. at $h/p=M/k$, where $M=0,1,2,...$ ($M<k$), the $k$-term does not contribute to the sum in (\ref{Supp:den}) and the corresponding resonance disappears in the response of the system. Note that the similar effect occurs in the case of the ordinary diffraction of light on a grating.

\begin{figure}[h]
    \centering
    \includegraphics[width=0.45\linewidth]{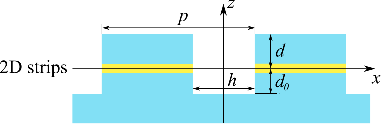}
    \caption{Schematic of a gap between the strips.}
    \label{STheor2}
\end{figure}

In principle, zeros of the expression~(\ref{Supp:den}), which correspond to the positions of the resonances, can be found numerically. However, in what follows we focus on the analytical analysis of the fundamental resonance only.

Before moving on, it is necessary to discuss the contribution of the dielectric permittivity in (\ref{Supp:den}), which is characterized by the value of $\varkappa_{k}$ defined in (\ref{Supp:E2}). Since 2D strips are fabricated by the etching, there is no dielectric in a gap, as shown in Fig.~\ref{STheor2}. The etching depth is $d+d_0$, where the additional depth $d_0$ is of the same order as $d=220$\,nm, i.e. $d_0$ is roughly from $100$ to $200$ nm. For small values of the ratio $h/p$, a significant part of the induced fields is concentrated in gaps between the stripes, where the dielectric is absent. This effect is not described by Eqs.~(\ref{Supp:resp}) and (\ref{Supp:den}) for the response of the system shown in Fig.~\ref{STheor}. That is why the contribution from $\varkappa_k$ in these equations should be adjusted. To describe the fundamental mode with low frequency $\omega^2 \ll \omega_p^2\varkappa_{k=1}$, we expect that the main contribution to the sum in Eqs.~(\ref{Supp:resp}) and (\ref{Supp:den}) will come from the first term with $k=1$. That is why, as an assessment we substitute $\varkappa_k$ with $\varkappa_{k=1}$ in~(\ref{Supp:resp}) and (\ref{Supp:den}). Note that in our system $\exp(-4 \pi d/p) \approx 0.9$, so $\varkappa_{k=1}$ is close to $\bar{\varkappa}=2\varepsilon/(\varepsilon+1)$, which corresponds simply to the average dielectric permittivity, as if 2D stripes were located just at the interface between the substrate and vacuum. As for higher modes, it seems that to estimate the frequency of mode with number $N$ one can substitute in (\ref{Supp:den}) $\varkappa_{k}$ with $\varkappa_{N}$.

We note that such a substitution of $\varkappa_k$ is likely inapplicable for the 'smallest' values of $h$, when electric fields are completely concentrated in the gaps, where the dielectric permittivity is equal to one. However, as will be shown below, even for the smallest experimental $h=1.3$\,$\mu$m, the contribution to the frequency from the interaction between strips is still not dominant.

Resonant frequencies are defined by the vanishing of the relation in (\ref{Supp:den}), which with the use the above mentioned assumptions and for the case of the fundamental resonance, becomes as follows:
\begin{equation}
\label{Supp:fund0}
    1+ 2\left(\frac{p}{h} \right)^2 \sum_{k=1}^{+\infty} \frac{1}{1-\omega_p^2 \varkappa_{k=1} k/\omega^2}\left(\frac{\sin(\pi k h/p)}{\pi k}\right)^2 =0.
\end{equation}

Since we consider the fundamental resonance we assume that the low-frequency condition $\omega^2 \ll \omega_p^2 \varkappa_{k=1}$ is satisfied and therefore we neglect the unity in the denominator of terms in the sum with $k \geq 2$, i.e. leave the unity only in the first term $k=1$. Then Eq.~(\ref{Supp:fund0}) appears as
\begin{equation}
\label{Supp:fund}
    \frac{h^2}{2p^2} +\frac{\Omega^4}{\Omega^2-1}\frac{\sin^2(\pi h /p)}{\pi^2} -\Omega^2 \sum_{k=1}^{+\infty} \frac{\sin^2(\pi k h /p)}{\pi^2 k^3}=0, \quad \text{where}\quad \Omega^2=\frac{\omega^2}{\omega_p^2 \varkappa_{k=1}}.  
\end{equation}
Then at the condition $h/p \ll 1$ the sum can be estimated as follows:
\begin{equation}
\label{Supp:main}
    \sum_{k=1}^{+\infty} \frac{\sin^2(\pi k h /p)}{\pi^2 k^3}\approx \left(\frac{h}{p}\right)^2 \left(\ln\frac{p}{2 \pi h}+\frac{3}{2}\right).
\end{equation}

Finally, multiplying Eq.~(\ref{Supp:fund}) by $\Omega^2-1$ and neglecting $\Omega^4$-terms one can find
\begin{equation}
\label{Supp:freq}
    \Omega^2=\frac{1}{2\left(\ln \left( \dfrac{p}{2 \pi h}\right) + 2\right)} \quad \text{or} \quad  
    \omega_1^2 =\frac{2 \pi n_s e^2 }{m^{\ast}  \varepsilon_1} \, \frac{\pi}{p} \, \frac{1}{\ln \left( \dfrac{p}{2 \pi h}\right) + 2},  \quad \text{where} \quad 
    \varepsilon_1=\frac{\varepsilon}{\varkappa_{k=1}}=\frac{\varepsilon}{1+ \frac{\varepsilon-1}{\varepsilon+1} e^{-4 \pi d/p}}.
\end{equation}

To qualitatively estimate the influence of the interaction between the strips on the resonance, let us compare the relation in (\ref{Supp:freq}) with the frequency of the fundamental plasmon mode in a single strip, situated just at the interface between dielectric medium with permittivity $\varepsilon$ and vacuum. This frequency appears as follows~\cite{Quinn:1986,Rudin:1997}:
\begin{equation}
\label{Supp:single_str}
    \omega_{single\,strip}\approx 0.87\sqrt{\frac{2 \pi n_s e^2 }{m^{\ast}  \bar{\varepsilon}} \frac{\pi}{w}},
\end{equation}
where $w$ is the width of a strip and $\bar{\varepsilon}=(\varepsilon+1)/2$. At $h \ll p$ we can put $w \approx p$ in the above equation and if one neglects $\ln(p/2\pi h)$ in Eq.~(\ref{Supp:freq}) then relations in (\ref{Supp:freq}) and (\ref{Supp:single_str}) become very similar, since $ \bar{\varepsilon}\approx \varepsilon_1$ and coefficients $1/\sqrt{2}\approx 0.7$ and $0.87$ are relatively close. That is why one can qualitatively think that $\ln(p/2\pi h)$ in the denominator of (\ref{Supp:freq}) describes the interaction of the charge density in different strips, while term '2' correspond to the interaction of charges inside a single strip. In the experiment under consideration even for the smallest $h=1.3$\,$\mu$m (at $p=25$\,$\mu$m) one finds $\ln(p/2\pi h)\approx 1.1$, which is of the same order as 2. This means that the interaction of charges in different strips and, correspondingly, the contribution of electric fields in gaps are not dominant compared to those inside a strip. 

The value of effective dielectric permittivity, $\varepsilon_1$, in (\ref{Supp:freq}) is in fact close to the average permittivity, $\bar{\varepsilon}=(\varepsilon+1)/2$; namely, for GaAs with $\varepsilon_{\rm GaAs}=12.8$ one can obtain $\varepsilon_1=7.2$ and $\bar{\varepsilon}=6.9$. The reason we retain the dependence on distance $d$ in Eq.~(\ref{Supp:freq}) is that it describes correctly the resonant frequency of plasmons in 2D strips embedded in infinite dielectric medium. Indeed the latter case corresponds to the formal limit $d \to \infty$, therefore $\varkappa_k=1$ for any $k$ in Eqs.~(\ref{Supp:resp}) and (\ref{Supp:den}) leading to $\varepsilon_1=\varepsilon$ in Eq.~(\ref{Supp:freq}).

However, to obtain a simpler and clearer expression, we replace $\varepsilon_1$ in Eq.~(\ref{Supp:freq}) with the average permittivity $\bar{\varepsilon}=(\varepsilon_{\rm GaAs}+1)/2$, i.\,e. take $d$ to be equal to zero. The resulting formula is given by Eq.~(2) of the main text and the corresponding resonant frequencies described by the simple formula are presented in Fig.~3 of the main text by solid red line. The remarkable agreement between the theory and experimental data takes place and, surprisingly, the agreement still remains even for $h/p \approx 0.5$, though Eq.~(\ref{Supp:freq}) was derived at $h/p \ll 1$.

Finally, we give an estimation of the influence of the substrate thickness, $D$, which is from $35$ to $45\, \mu$m, on the resonant frequency. The influence manifests itself in the same way as that of the finiteness of distance $d$ in~(\ref{Supp:freq}), namely, through the factor $\exp(-4 \pi D/p)\approx 10^{-7}$ (at $D=35\, \mu$m and $p=25\,\mu$m), which is negligibly small. Thus, the substrate thickness can be considered being infinite, though $D$ and $p$ are of the same order. 

\newpage

\section{\textrm{III}. Analysis of the higher-order plasmon modes in the case of a small inter-strip gap}

To elucidate the origin of the observed superlattice plasmon mode, it is useful to examine the harmonic plasmon modes excited in the plasmonic metasurface. For simplicity, we consider the dispersion equation~(\ref{Supp:fund0}) in the limiting case where the strips are located directly at the substrate--vacuum interface:
\begin{equation}
\label{Supp:fund_av}
    1+ 2 \sum_{k=1}^{+\infty} \frac{1}{1-\omega_{p,\mathrm{av}}^2 k/\omega^2}
    \left(\frac{\sin(\pi k h/p)}{\pi k h/p}\right)^2 =0,
    \qquad \text{where} \qquad
    \omega_{p,\mathrm{av}}^2 = \frac{2 \pi n_s e^2 }{m^{\ast}\bar{\varepsilon}} \, \frac{2\pi}{p}.
\end{equation}
Here, $\bar{\varepsilon}=(\varepsilon_{\rm GaAs}+1)/2$ denotes the effective (average) dielectric permittivity.

\begin{figure}[h]
    \centering
    \includegraphics[width=0.6\linewidth]{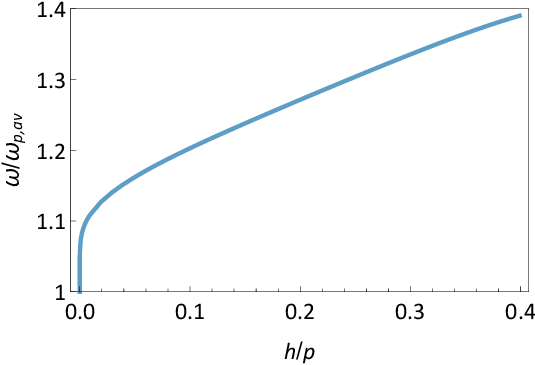}
    \caption{Frequency of the first excited plasmon mode ($N=1$) as a function of the normalized gap width $h/p$, obtained from Eq.~(\ref{Supp:fund_av}). In the limit $h\to 0$, the frequency approaches $\omega_{p,\mathrm{av}}$, corresponding to the ordinary 2D plasmon with the wave-vector $2 \pi/p$.}
    \label{Suppsmallh}
\end{figure}

We now examine the limit $h\to 0$. In this limit, the series in Eq.~(\ref{Supp:fund_av}) diverges logarithmically, $\sim \ln h$. Hence, Eq.~(\ref{Supp:fund_av}) can be satisfied in two qualitatively different ways. The first option is to let $\omega \to 0$ as $h\to 0$. This low-frequency superlattice mode is analyzed in the previous section. The second option is realized when $\omega$ approaches a pole of one of the terms in the series, when
\begin{equation}
\label{Supp:poles}
\omega=\omega_{p,\mathrm{av}} \, \sqrt{N}, \qquad N=1,2,\ldots,
\end{equation}
so that the denominator of the $N$th term in Eq.~(\ref{Supp:fund_av}) vanishes. To demonstrate this explicitly, we single out the $N$th term and rewrite Eq.~(\ref{Supp:fund_av}) as
\begin{equation}
\label{Supp:fund_avN}
1+\frac{2}{1-\omega_{p,\mathrm{av}}^{2}N/\omega^{2}}
+2\sum_{k=1,\,k\neq N}^{\infty}
\frac{1}{1-\omega_{p,\mathrm{av}}^{2}k/\omega^{2}}
\left(\frac{\sin(\pi k h/p)}{\pi k h/p}\right)^{2}=0.
\end{equation}
Here, in the second term we have already taken the limit $h\to 0$ for the factor
$\left[\sin(\pi N h/p)/(\pi N h/p)\right]^2\to 1$. The remaining sum in Eq.~(\ref{Supp:fund_avN}) still diverges as $h\to 0$. Therefore, for Eq.~(\ref{Supp:fund_avN}) to hold at a finite frequency, the isolated second term must also become singular, which yields Eq.~(\ref{Supp:poles}). The discrete frequencies~(\ref{Supp:poles}) correspond to the conventional 2D plasmon dispersion $\omega\propto\sqrt{q}$ [see Eq.~(1) from the main text] evaluated at the reciprocal-lattice wave vectors $q=2\pi N/p$. The dependence of the $N=1$ mode frequency on $h/p$, obtained from Eq.~(\ref{Supp:fund_av}), is shown in Fig.~\ref{Suppsmallh}.

\begin{figure}[h]
    \centering
    \includegraphics[width=0.9\linewidth]{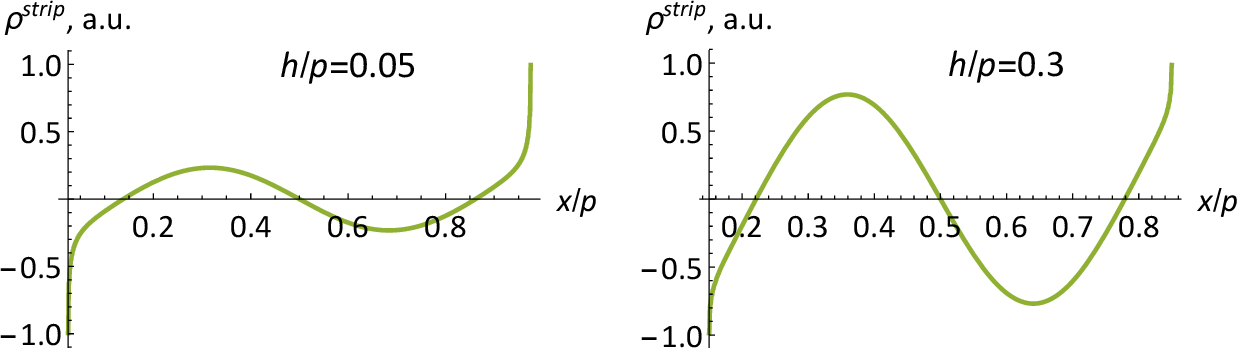}
    \caption{Charge density in a strip, $\rho^{strip}$, in the first exited plasmon mode, $N=1$, for $h/p=0.05$ and $0.3$ is shown as a function of coordinate $x/p$.}
    \label{SuppC}
\end{figure}

Also, using Eq.~(\ref{Supp:E2}) the charge density in a strip, $\rho^{\rm strip}(x)$, can be determined, as $\rho^{\rm strip}(x) \propto \partial_x E_x(x)$. The distribution of the charge density for $N=1$ mode is presented in Fig.~\ref{SuppC} for $h/p=0.05$ and $h/p=0.3$.

\newpage

\section{\textrm{IV}. The influence of constant perpendicular magnetic field}
Consider the system shown in Fig.~\ref{STheor} placed in the constant magnetic field ${\bf B}$ directed along $z$-axis. Magnetic field leads to the appearance of the current density along the strips, $j_y$, and, in general, the excitation of transverse electric (TE) modes, which possess the remaining components of the fields ($H_x$, $E_y$, $H_z$). Note also, that the emergence of $E_y$ in the system in magnetic field corresponds to a kind of the Faraday effect.

To describe TE-modes we derive the relation for $E_y^k(z)$ and $j_y^{2D,k}$, which is the analog of Eq.~(\ref{Supp:waveX}). We exclude from Eqs.~(\ref{Supp:Maxeq}) fields $H_{x,z}^k(z)$
\begin{equation}
    H_x^k=-\frac{c}{i\omega}\partial_z E_y^k, \quad
    H_z^k=\frac{cq}{\omega} E_y^k, 
\end{equation}
and find the following:
\begin{equation}
\label{Supp:waveY}
    \partial_z^2 E_y^k(z) -\left( q^2-\frac{\omega^2}{c^2}\varepsilon(z)\right)E_y^k(z) =-\frac{4 \pi i \omega}{c^2} j_y^{2D,k}\delta(z).
\end{equation}

Now, we consider the scattering problem for $E_y^k(z)$, consequently, we look the solution in the form
\begin{equation}
\label{Supp:formY}
E_y^k(z)=
\begin{cases}
     C_{y,1}^k e^{-\beta^0_k z}, & z\geq d; \\
     C_{y,2}^k e^{\beta_k z}+ C_{y,3}^k e^{-\beta_k z}, & d>z>0; \\
     C_{y,4}^k e^{\beta_k z}, & z\le 0;
\end{cases}
\end{equation}
with notations discussed after Eq.~(\ref{Supp:formX}). The first boundary condition is the continuity of $E_y^k(z)$. The other can be found via the integration of Eq.~(\ref{Supp:waveY}) near $z=0$ and $z=d$ points and, consequently, one can find:
\begin{equation}
\label{Supp:BCy}
    \partial_z E_y^k(z) |^{z=+0}_{z=-0}= -\frac{4 \pi i \omega}{c^2 } j_y^{2D, k} \quad \text{and} \quad  \partial_z E_y^k(z) |^{z=d+0}_{z=d-0}=0.
\end{equation}

Finally, one achieves
\begin{equation}
\label{Supp:EqsEy}
    E_y^k(z=0)=\frac{2 \pi i\omega}{\beta_k c^2} \left(1+\frac{\beta_k^0-\beta_k}{\beta_k^0+\beta_k} e^{-2\beta_k d} \right) j_y^{2D,k},
\end{equation}
which together with Eq.~(\ref{Supp:eqsE}) and the Ohm's law form a closed set of equations that describes the response of the system.

The local Ohm's law appears as follows: 
\begin{equation}
\begin{pmatrix}
     &j_x^{2D}(x) \\
     &j_y^{2D}(x) 
\end{pmatrix}
=
\begin{pmatrix}
\sigma_{xx} & \sigma_{xy}\\
\sigma_{yx} & \sigma_{yy}
\end{pmatrix}
\begin{pmatrix}
     &E_x(x,z=0) \\
     &E_y(x,z=0) 
\end{pmatrix},
\end{equation}
where $x$ lies inside a strip, off-diagonal (Hall) components $\sigma_{xy}$ and $\sigma_{yx}$ of the conductivity tensor arise due to nonzero magnetic field $\bf{B}$ applied to the system and the usual relations $\sigma_{xy}=-\sigma_{yx}$ and $\sigma_{xx}=\sigma_{yy}$ take place.

In what follows, we use the Drude model for conductivities, which appears as:
\begin{equation}
    \sigma_{xx}=\frac{e^2 n_s}{m^{\ast}}\frac{-i\omega +1/\tau}{(-i\omega +1/\tau)^2+\omega_c^2} \quad \text{and} \quad \sigma_{xy}=\frac{e^2 n_s}{m^{\ast}}\frac{-\omega_c}{(-i\omega +1/\tau)^2+\omega_c^2}, \quad \text{where}\,\,\, \omega_c=\frac{eB}{cm^{\ast}}
\end{equation}
is the electron cyclotron frequency (here $e>0$ is the elementary charge).

Now, we introduce the quasi-electrostatic approximation, which was discussed after Eq.~(\ref{Supp:Efin}), and take $\beta_k=\beta_k^0=q_k$ for $k\geq 1$. Then one can obtain from Eq.~(\ref{Supp:EqsEy})
\begin{equation}
\label{Supp:Eyquas}
     E_y^{k=0}(z=0)=-\frac{2 \pi }{c}\frac{2}{1+\sqrt{\varepsilon}}j_y^{2D,k=0} \quad \text{and}  \quad E_y^k(z=0)=\frac{2 \pi i\omega}{q_k c^2} j_y^{2D,k} \quad \text{for}\,\, k \neq 0,
\end{equation}
where for $E_y^{k=0}$ we take into account that $ \omega d \sqrt{\varepsilon}\ll c$. 

To proceed we find out how a quasi-electrostatic limit looks like. The 'pure' quasi-electrostatic regime corresponds to the formal limit $c \to \infty$ and, consistently, the right-hand sides in (\ref{Supp:Eyquas}) vanish. Thus, in the 'pure' quasi-static limit $E_y(z=0)$ simply does not appear, which means that the influence of magnetic field $ {\bf B}$ is reduced to the elementary replacement of $\sigma$ with the longitudinal conductivity $\sigma_{xx}$ in Eq.~(\ref{Supp:jx2}) and, respectively, Eq.~(\ref{Supp:resp}). This finally leads to the emergence of the relation $\omega^2-\omega_c^2$ instead of $\omega^2$ in subsequent formulas including Eq.~(\ref{Supp:freq}). Thus, the frequencies of plasmon modes in nonzero magnetic field, $\omega(B \neq 0)$, can be found as $\omega^2(B \neq 0)=\omega^2(B=0)+\omega_c^2$~\cite{Theis:1977}.

Now, let us estimate the amplitude of $E_y(z=0)$ compared to $E_x(z=0)$ using Eqs.~(\ref{Supp:Eyquas}). In the case of $k=0$ we can roughly evaluate 
\begin{equation}
    E_y^{k=0}(z=0) \approx \frac{2 \pi \sigma_{xy}}{c}\frac{2}{1+\sqrt{\varepsilon}}\frac{w}{p} E_x^{av}, 
\end{equation}
where $E_x^{av}$ is the average value of the field across a strip and the 'self-contribution' $\propto \sigma_{xx}E_y$ in the right-hand side of the above relation is neglected. 

The thing is that the parameter connecting $E_y^{k=0}$ and $E_x^{av}$, generally speaking, is not small. For example, at the cyclotron resonance, $\omega=\omega_c$, on can find
\begin{equation}
    \left| \frac{2 \pi \sigma_{xy}(\omega=\omega_c)}{c}\right|  \frac{2}{1+\sqrt{\varepsilon}} \approx 0.61,
\end{equation}
where we take $n_s=9.3\times 10^{11}$~cm$^{-2}$, $m^{\ast} = 0.071 m_0$, $\tau =4$\,ps, and $\varepsilon=12.8$, which corresponds to the experiment under consideration. 

That is why, we find the above parameter at the frequency of the plasmon mode in quasi-electrostatic regime, i.e. introduce in $\sigma_{xy}(\omega)$ the value $\omega=\sqrt{\omega_1^2+\omega_c^2}$, where $\omega_1$ is defined by Eq.~(\ref{Supp:freq}). Then one can find
\begin{equation}
  R=  \left| \frac{2 \pi \sigma_{xy}\left(\omega=\sqrt{\omega_1^2+\omega_c^2}\right)}{c}\right|  \frac{2}{1+\sqrt{\varepsilon}} \frac{w}{p}=\frac{2\varepsilon_1\left(\ln(p/2 \pi h )+2\right)}{1+\sqrt{\varepsilon}} \frac{\omega_c}{\pi c}w.
\end{equation}
The parameter grows linearly with the increase of the cyclotron frequency and at the smallest experimental $h=1.3\,\mu$m and $B=5$ kG one achieves $R\approx 0.3$. Thus, roughly speaking, at $B<5$ kG the electric field $E_y^{k=0}$ can be neglected, while at $B>10$ kG the parameter $R$ becomes of the order of one and the emergence of $E_y^{k=0}$ should be taken into account. However, the consideration of its influence on the plasmon resonances is beyond the scope of this paper.

As for $y$-components of the electric field with $k\neq 0$ in Eq.~(\ref{Supp:Eyquas}), the ratio of $E_y^k(z=0)$ to a corresponding of $x$-component of the electric field is qualitatively governed by the parameter
\begin{equation}
\label{Supp:Eynz}
   \frac{2 \pi i\omega}{q_k c^2}\sigma_{yx} \frac{w}{p}=\frac{2 \pi \sigma_{yx}}{c}i \frac{w}{\lambda_0 k}, 
\end{equation}
where we neglect the contribution $\propto \sigma_{xx}E_y$, while deriving the above relation from Eq.~(\ref{Supp:Eyquas}), and $\lambda_0$ is the wavelength of light in vacuum. Important, that the additional parameter, $w/ \lambda_0$, appears in comparison to the case of $k=0$. For typical frequency of $200$ GHz we have $\lambda_0=1.5$\,mm, so $w/ \lambda_0 \approx p/\lambda_0=1.6 \cdot 10^{-2}$. That is why, the parameter in~(\ref{Supp:Eynz}) is much less that one even if $2 \pi \sigma_{yx}/c$ is of the order of one; consequently, the appearance of $E_y^{k \neq 0}$ can be neglected.

\section{\textrm{V}. Additional experimental data for superlattices with different periods}

\begin{figure}[h]
    \centering
    \includegraphics[width=0.8\linewidth]{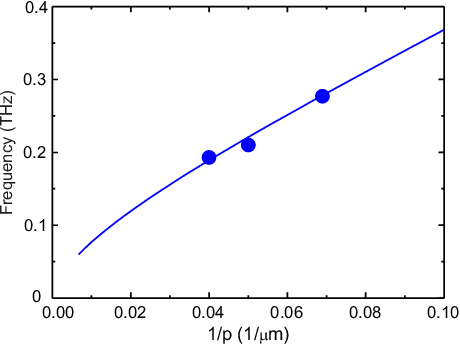}
    \caption{Superlattice plasmon frequency versus inverse period ($1/p$) for samples with a fixed gap between 2DES strips ($h=2$~\textmu{}m) and superlattice periods $p=25, 20$ and $14.5$~\textmu{}m. The solid line shows the theoretical prediction from Eq.~(\ref{NewPlasmon}). The data agree well with theory, supporting the proposed model across varying superlattice periods.}
    \label{S3}
\end{figure}

Experiments analogous to Fig.~3 of the main text were performed on a series of samples with a fixed gap between the 2DES strips, ($h=2$~\textmu{}m), and superlattice periods $p=25, 20$ and $14.5$~\textmu{}m. Figure~\ref{S3} plots the superlattice plasmon frequency as a function of the inverse period, $1/p$. The solid line represents the theoretical prediction:
\begin{equation}
    \omega_{sp} =\sqrt{\frac{n_s e^2 }{2 \, m^{\ast} \varepsilon_0  \varepsilon_{\rm eff}} \, \frac{\pi}{p} \, \frac{1}{\ln \left( \dfrac{p}{2 \pi h}\right) + 2}},  
\label{NewPlasmon}
\end{equation}
There is a good agreement between the theory and experimental data, supporting the validity of the proposed theory for varying superlattice periods.

\newpage

\bibliography{main}